\documentclass{kapproc} 
\usepackage{graphicx}

  \usepackage{procps}

\normallatexbib

\begin{document}

\articletitle[$\Sigma(1385)$ Resonance Studies with STAR at
$\sqrt{s_{NN}}= 200$ GeV ] {$\Sigma(1385)$ Resonance Studies
\\with STAR at $\sqrt{s_{NN}}= 200$ GeV}

\author{Sevil Salur for STAR Collaboration}

\affil{Yale University, Physics Department,\\ 272 Whitney Ave. New
Haven, CT} \email{sevil.salur@yale.edu}


\begin{abstract}
In p+p, d+Au, and Au+Au collisions at $\sqrt{s_{NN}}=200$ GeV with
the STAR detector at RHIC, $\Sigma(1385)\rightarrow\Lambda+\pi$
were measured using two techniques; three-particle mixing and a
hybrid mixing technique. We present results from both of these
methods and compare the invariant mass spectra and the
backgrounds.
\end{abstract}

\begin{keywords}
Resonances, Relativistic Heavy Ion Collisions
\end{keywords}

\section*{Introduction}
Due to the very short lifetime ($\sim$ few fm/c) of resonances a
large fraction of the decays occur inside the reaction zone in the
most central relativistic heavy ion collisions. These resonances
have a high probability of not being reconstructed as their
daughter particles are likely to re-scatter in the medium.
Resonances with higher transverse momentum are more likely to be
reconstructed because of their longer relative lifetimes. This
means they are more likely to decay outside of the medium and
hence their daughter particles would interact less with the
medium\cite{marcus}. This may give rise to effects such as changes
in the physical properties of the resonances, width broadening and
mass shifts, or changes in the transverse momentum ($p_{T}$)
spectra due to re-scattering of the decay particles. Hence the
study of resonances provides an additional tool to determine the
hadronic expansion time between chemical and thermal freeze-out.
Due to its strange quark content and high mass, $\Sigma(1385)$ may
also give additional information about strangeness enhancement,
one of the possible signatures of the Quark Gluon Plasma
\cite{raf:1}.

There are also several models currently available that attempt to
describe hadronic matter with resonances. One is based on the
thermal model and argues that the direct measurement of resonances
can probe both the hadronization temperature and the lifetime of
the interacting hadron gas phase \cite{raf:2}. In this model, the
ratios of resonances to lighter particles with identical number of
valance quarks are sensitive to the freeze-out temperature because
all of the chemical dependence cancels out in the ratio.

\section{Analysis and Particle Identification}

Charged daughter particles are identified by the momentum and the
energy loss per unit length (dE/dx) measured in the Time
Projection Chamber of STAR (TPC)\cite{Star}. For long-lived
particles ($c\tau\sim$ few cm\c) such as the $K^{0}_{s}$,
$\Lambda$, and $\Xi$, the decay vertex topology is used for
identification. This technique cannot be used for resonances due
to their very short lifetimes ($c\tau_{\Sigma(1385)}=5 fm/c$).
However an alternative method, called the mixing technique, can be
used and also applied to identify other short-lived particles such
as pentaquarks and dibaryons.

The decay channel that we investigate is
$\Sigma(1385)\rightarrow\Lambda+\pi$. We use two methods to
identify $\Lambda$'s. In the first method, the three particle
mixing technique (TPM), every $\pi$ is combined with every $p$ to
produce a $\Lambda$ candidate. Then the $\Lambda$ candidate is
combined with all remaining $\pi$ to get $\Sigma(1385)$. In the
second method, the hybrid mixing technique (HM), $\Sigma(1385)$'s
are identified by combining topologically reconstructed
$\Lambda$'s with $\pi$'s. In both techniques the background is
described by combining $\pi$ 's from one event with the
$\Lambda$'s from another event.
\section{Status of Current Studies}
$\Sigma(1385)\rightarrow \Lambda+\pi_{bachelor}$ is identified in
p+p, d+Au, and Au+Au collisions with the mixing techniques. Figure
1 is the invariant mass spectra of the $\Sigma$(1385) for the
various nuclear systems studied. With the TPM technique, in the
lower kinematic limit of the invariant mass spectrum a background
structure appears (Fig 1-a). This initial structure increases in
the d+Au collisions and totally dominates the spectrum in Au+Au
collisions, due to an increase in background combinatorial
statistics. Monte Carlo studies show that a partial contribution
of the background structure comes from the misidentification of
the $\Lambda$'s with the $\pi_{bachelor}$ and misidentification of
K's as $\pi$'s. A more detailed study with the simulations is
needed to find the ratio of each contribution and the other
possible sources. The HM technique can be used to identify
$\Sigma$(1385) with less efficiency but with a cleaner background
(Fig 1-b,c,d). Since $\Lambda$'s can be identified more cleanly
with the decay topology technique, the initial structure
disappears. Antiparticle to particle ratios of $\Xi$ and
$\Sigma$(1385) are observed as $0.89\pm0.04$ and $0.90\pm0.07$
respectively in p+p collisions. These values are consistent with a
net baryon-free collision environment.

With available events, it is possible to study the $p_{T}$ spectra
of the $\Sigma$(1385) for all systems. In order to correct for
STAR's efficiency and acceptance, a study using Monte Carlo tracks
embedded into real events is necessary.
\begin{figure}[h!]
\centering
\includegraphics[height=7cm,clip]{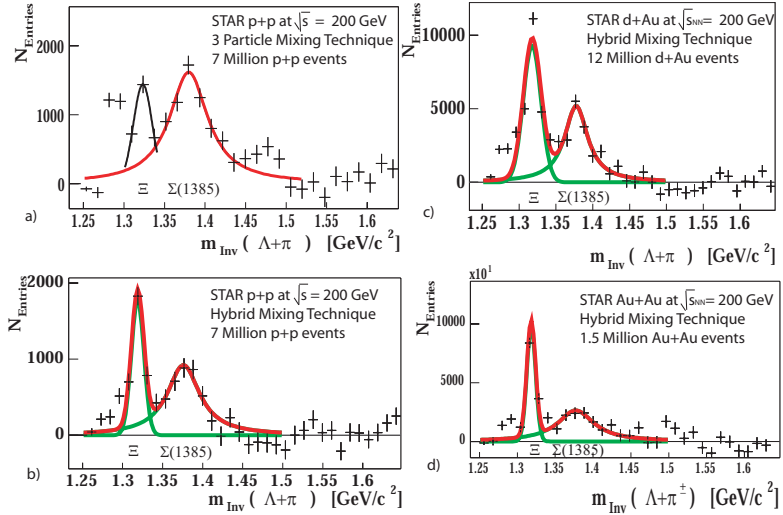}
 \caption{Invariant mass spectra of $\Xi$ and $\Sigma$(1385) after
 mixed-event background subtraction:
 (a) Invariant mass spectrum of $\Sigma$(1385) in p+p using the three particle mixing
 technique. The $\Xi$ has the same decay channel as the $\Sigma$(1385). The $\Xi$ is fit with a gaussian
distribution and the $\Sigma$(1385) is fit with a Breit-Wigner.
 (b) Invariant mass spectrum in p+p using the hybrid mixing technique. With respect to the TPM technique
 the structure at the low $M_{inv}$ in the background disappears with cleanly identified $\Lambda$'s.
 (c) Invariant mass spectrum in d+Au using the hybrid mixing technique.
 (d) Invariant mass spectrum in Au+Au using the hybrid mixing technique.}
\end{figure}
\section{Summary}Using the mixing technique $\Sigma$(1385) and $\Xi$ are constructed in $\sqrt{s_{NN}} = 200$ GeV
p+p, d+Au, and Au+Au data. Raw yields and ratios are presented.
Corrections to these values are being investigated and will be
resolved in the very near future. Fireball temperature of the
collision can be studied using the $p_{T}$ distributions after the
corrections. Studying Au+Au collisions is essential to understand
possible rescattering and medium effects on resonances. It is
possible to study pentaquarks with STAR and with the upcoming run
starting this January 2004 we will continue our investigation of
exotic particles by using the STAR Silicon Vertex Tracker (and
eventually a new Time of Flight detector system).

\end{document}